\documentstyle[twocolumn,prb,aps,epsf]{revtex}
\def\rect#1#2{{\vcenter{\vbox{\hrule height.3pt
            \hbox{\vrule width.3pt height#2truecm \kern#1truecm
            \vrule width.3pt}
            \hrule height.3pt}}}}
\def\square{\rect{0.15}{0.15}}
\def\blacksquare{\vrule height 0.15truecm width 0.15 truecm depth -0.1ex}
\def\inseps#1#2{\def\epsfsize##1##2{#2##1} \centerline{\epsfbox{#1}}}
\begin{document}
\draft
\input{psfig}
\title{Anisotropic dynamical scaling in a spin model with competing 
interactions}
\author{E.N.M. Cirillo, G. Gonnella, 
and S. Stramaglia [\onlinecite{ENZO}]}
\address{Dipartimento di Fisica dell'Universit\`a di Bari and
Istituto Nazionale di Fisica Nucleare, Sezione di Bari, via Amendola
173, 70126 Bari, Italy. E\_mail: cirillo@ba.infn.it, gonnella@ba.infn.it,
stramaglia@axpba5.ba.infn.it}
\date{\today}
\maketitle
\begin{abstract}
Results are presented for the kinetics of domain growth of a
two-dimensional Ising spin model with competing interactions
quenched from a disordered to a striped
phase. The domain growth exponent are $\beta=1/2$ and 
$\beta=1/3$ for single--spin--flip and spin--exchange dynamics,
as found in previous simulations.
However the correlation functions measured in the direction 
parallel and transversal to the stripes are different as suggested 
by the existence of different interface energies 
between the ground states of the model.
In the case of single--spin--flip dynamics an anisotropic version
of the Ohta-Jasnow-Kawasaki theory for the pair scaling function 
can be used to fit our data.
\end{abstract}


The kinetics of phase separation as  systems with competing interactions
are quenched below their ordering temperature  is the subject 
of much current research in a variety of fields in physics
and other sciences [\onlinecite{BBB}].
Sometimes the equilibrium
configuration of these systems is a lamellar structure with the
ordered phases arranged in stripes. Our aim here is to study
the growth of striped phases in a generalization of the
2D Ising model. In particular, we will focus our attention on 
dynamical properties of the correlation functions 
in the directions parallel and transversal
to the stripes.
 
In the simplest cases the late time ordering process
is characterized by a single time dependent length
$R(t) \sim t^{\beta}$ representing the average domain
size [\onlinecite{B94}]. 
This implies a particular behavior for the correlation
function: if $\varphi (\vec x,t)$ is the ordering field the equal time
correlation $C(\vec r, t) \equiv 
<\varphi (\vec x,t)\varphi (\vec x + \vec r,t)>$ has the scaling form 
$C(\vec r, t) = f(r/R(t))$ where f(z) is a scaling function.
Our results for the striped phase
show the same growth exponent in any direction relative
to the stripes and an explicit dependence 
of the {\em different} longitudinal and  transversal
scaling functions on the microscopic details of the system. 
We will try to understand this difference and, 
in the   case of single--spin--flip dynamics, we interpret our data by 
 a generalization to anisotropic cases
of the Ohta-Jasnow-Kawasaki theory [\onlinecite{OJK}] for the scaling function.
The anisotropic
OJK theory is then tested on the 
Ising model with different nearest-neighbor couplings for the two
square lattice directions.

Lamellar phases appear in many physical systems. Examples are
diblock copolymer melts [\onlinecite{B91}],
surfactant--oil--water mixtures [\onlinecite{GS94}],
dipolar fluids with long-range Coulombic
interactions[\onlinecite{SD93}]. Differently than in these
systems,   the striped phase of this paper is not characterized by
any mesoscopic length - the width of the lamellae is always
one lattice spacing in the model here considered.
Neverthless our results may be relevant for systems
with more realistic lamellar phases. Moreover,
from a more theoretical point of view, 
it is interesting to have an explicit example of how the 
scaled
correlation functions can depend on the details of the system,
also in relation with  recent discussions on this theme
[\onlinecite{Rut}].

The model we consider is the well-known Ising version
of the bidimensional isotropic eight vertex model [\onlinecite{Bax}], 
with hamiltonian $H$ given by
\begin{equation}
-\beta H = J_1 \sum_{<ij>} s_i s_j + J_2\sum_{<<ij>>} s_i s_j
          +J_3 \sum _{[i,j,k,l]} s_i s_j s_k s_l,
\label{eq:ham}
\end{equation}
where $s_i$ are Ising spins on a bidimensional square lattice and the sums 
are respectively on nearest neighbor pairs of spins, next to the nearest 
neighbor pairs and plaquettes. Periodic
boundary conditions are always assumed.
At $J_2 < 0, |J_1| < 2 |J_2|$ and $J_3$ small
a critical line separates the paramagnetic phase
from a region where four phases corresponding to ground states
with alternate plus and minus columns or rows coexist [\onlinecite{Bax}].
Sudden quenches of the system from a completely disordered
initial configuration to the stripe-ordered phases will be 
studied by numerical simulations both at 
zero and finite temperature, for different values of the parameters,
with heat--bath single--spin--flip  [\onlinecite{Gla}] 
and  spin--exchange dynamics [\onlinecite{Kawa}].

Before presenting our results it is useful to summarize the 
known behavior of  system (1) during 
the phase separation process. 
First consider quenching in the ferromagnetic phase in both $d=2$
and $d=3$. When  all the couplings are ferromagnetic
the growth exponent is $\beta=1/2$ or $\beta=1/3$
[\onlinecite{B94},\onlinecite{Leb},\onlinecite{Bark}] 
corresponding  respectively to single--spin--flip and  spin--exchange dynamics. 
The situation becomes different for a weak
antiferromagnetic  coupling $J_2$ when,
still in the ferromagnetic phase, energy barriers
oppose to the coarsening of the domains and the system does not
relax to equilibrium if the temperature is zero [\onlinecite{Sh}]. 
In $d=3$ the energy
barriers are proportional to the linear dimension of domains and a 
 logarithmic growth is expected [\onlinecite{Sh}]. Quenching in the 
striped phase in $d=2$ have  been already studied in [\onlinecite{Bind}].
The simulations of [\onlinecite{Bind}] show that also in the striped
phase the average size of domains grows as $R(t) \sim t^{\beta}$ with 
$\beta=1/2$ or $\beta=1/3$ for single--spin--flip and spin--exchange
dynamics.
Motivated by the idea of studying asymptotic correlations
in lamellar phases we have analyzed again the growth and the
dynamical scaling in the striped phase of model (1).

\begin{figure}
\vskip -10mm
\inseps{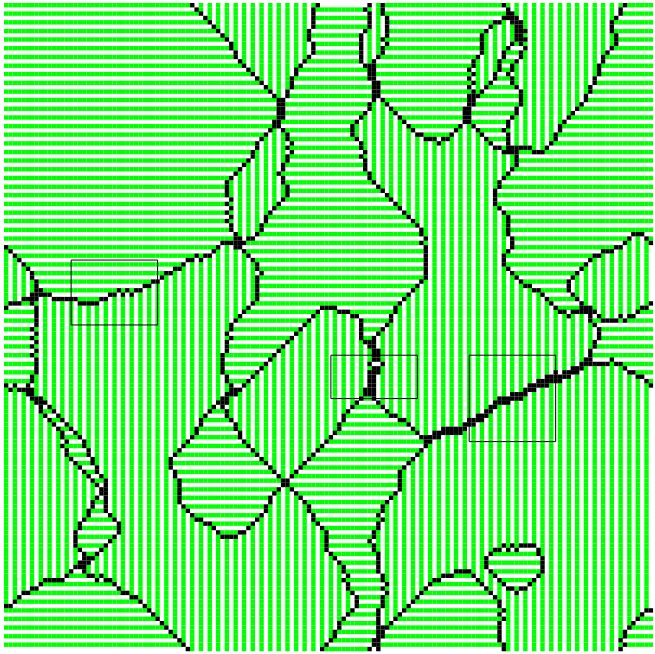}{0.55}
\vskip -10mm
\caption{Typical configuration of model (\ref{eq:ham}) in the
scaling regime.
Grey and white squares represent respectively plus and minus spins;
black squares represent the interface sites.
The picture refers to a $100\times 100$ square lattice, at zero temperature,
with parameters $J_1=0.1$, $J_2=-1.0$ and $J_3=0$,
after $150$ MCS (Monte Carlo Steps per site). The three boxes
put in evidence different kinds of interfaces.}
\end{figure}

A typical configuration of the system during the evolution
after a quench at $T=0$ is given in Fig.1. The configuration is a patchwork
of vertical and horizontal striped domains. Simulations show that
in the case of single--spin--flip dynamics 
there are no energy barriers and  the system separates also at 
zero temperature. To monitor the domain growth we evaluated the amount 
of interfaces present in the system. To decide if a given site belongs
or not to an interface, we compare the configuration of the system
in a neighborhood of the site with four given patterns corresponding to the 
four ground states of the model. Then it is possible to define a distance 
between each pattern and the local configuration of the system.
If this distance is greater than some threshold we say the site
belongs to an interface, otherwise it is a part of some domain 
[\onlinecite{nota}].
Our results have been shown not to depend on the values of 
the threshold and of the size of the patterns to be compared.
Interfaces identified in this way are 
shown in Fig. 1.
In $d=2$ the total length of interfaces per unit volume
$L$ scales as the inverse
of the average size of domains. The results of 
[\onlinecite{Bind}] are confirmed by our 
simulations: we find $L \sim t^{-0.5}$ in the case
of single--spin--flip dynamics and $L \sim t^{-1/3}$
in the spin--exchange dynamics.
The same conclusions have been obtained also 
monitoring the shrinking of a $N \times N$ square domain
of one phase immersed in a sea of the three other phases. In this case
the dipendence on $N$ of the time of shrinking can be used 
to evaluate the growth exponent.

We now turn to consider the correlation properties of growing domains in 
the scaling regime. We introduce 
the equal time 
{\it longitudinal} and {\it transversal} correlation functions
respectively defined as 
\begin{equation}
\begin{array}{lll}
C_{\ell} (r,t) &=&\big< s\left( i,j\right) s\left( i+\epsilon (i,j) r\; ,\; 
j+(1-\epsilon(i,j)
) r\right) \big>\\
&&\\
C_t (r,t) &=&\big< {(-1)}^r s\left( i,j\right) s\left( i+(1-\epsilon (i,j)) r
\; ,\;
j+ \epsilon(i,j)
    r\right) \big>\\
\end{array}
\label{eq:cl}
\end{equation}
where $\epsilon (i,j)$ is one (zero) if site $(i,j)$ belongs to a horizontal 
(vertical) domain and the average is performed over all the sites not 
belonging to interfaces at time $t$ and
over different stories of the system.
These correlations have the property that for a fixed
$(i,j)$ they become zero outside a given domain. 
The scaling behavior
of $C_{\ell}$ and $C_t$  for a particular case with single--spin--flip dynamics
is shown in Fig.2
 where data taken at different times
have been plotted in terms of the scaling variable $z=r/\sqrt{t}$.
The parameters of the simulations of Fig.2 were $J_1=0.1,
J_2=-1, J_3=0$.
We see that the longitudinal and the transversal
scaling functions $f_{\ell}$ and $f_t$ are different,
which is a general feature
turning out from our simulations [\onlinecite{nota1}].
Actually the difference between $f_{\ell}$ and $f_t$
is very tiny in the case of spin--exchange dynamics,
but always understandable on the basis of the arguments given below.
In the following we will show results only for simulations
with single--spin--flip dynamics.

To explain the observed difference
between $f_{\ell}$ and $f_t$ a
simple argument can help:  a ferromagnetic coupling
$J_1$ would favour longitudinal with respect to
transversal correlations while the reverse is true
when $J_1$ is negative. This is what happens, indeed.
The results of simulations at $J_1=-0.1,
J_2=-1, J_3=0$ are the same as those of Fig. 2 but with
the role of $C_{\ell}$ and $C_t$ reversed.
Not only, fixed $J_2$ and $J_3$, $C_{\ell}$ is always greater
 than $C_t$ for $J_1>0$ while  $C_{\ell} < C_t$ for $J_1<0$,
but the behavior of  $C_{\ell}$ and $C_t$ is very symmetric
with respect to the change of the sign of $J_1$.
An heuristic argument, based on the existence of interfaces with different
surface tensions between the 4 different domains, 
can explain this symmetry. For simplicity consider 
interfaces parallel to the lattice directions and $J_3=0$.
There are three types of these interfaces all depicted
in the boxes in Fig. 1. On the left of the picture a boundary 
between a vertical and a horizontal domain is marked;
this interface reduces both longitudinal and transversal
correlations and it does not matter for our argument.
The situation is different for the two other kind of interfaces
marked in Fig.1: the interface in the middle reduces only transversal correlations
while the one on the right reduces only longitudinal correlations
not influencing the transversal ones.
The T=0 excess energy of the interface in the middle is 
$2 J_2 - J_1 $ while the excess energy for the interface on the right
is $2 J_2 + J_1 $. Since the excess
energy is the driving force for the phase separation process,
if we assume that the role of these two interfaces
is the same during this process,
we can expect a symmetric
behavior of the longitudinal and transversal correlation functions
with respect to the change of sign of $J_1$.

\begin{figure}
\vskip -7mm
\inseps{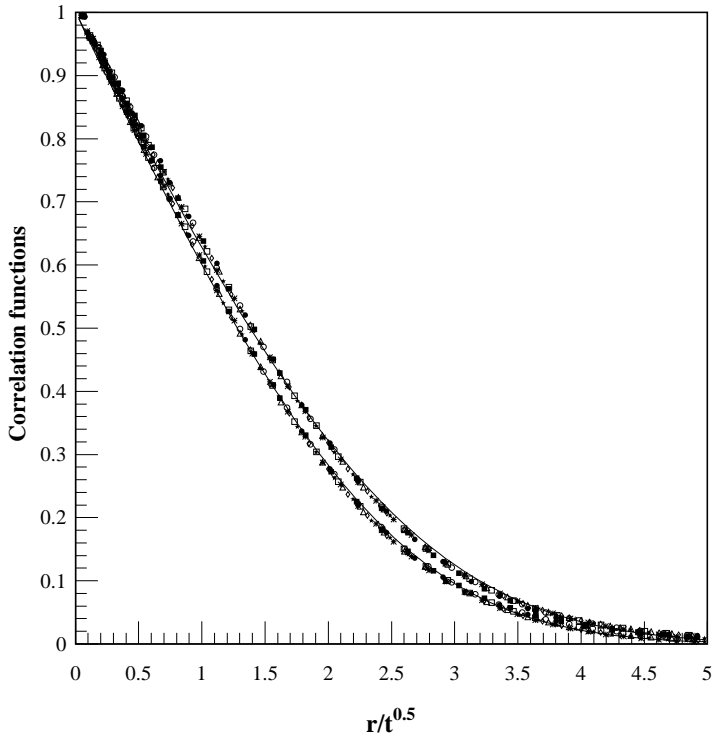}{0.6}
\vskip -18mm
\caption{Scaling functions $f_{\ell}(z)$ and
$f_t(z)$, $z=r/t^{0.5}$ in the case of single--spin--flip
dynamics
with $T=0$, $J_1=0.1$, $J_2=-1$ and $J_3=0$.
We averaged over $50$ different histories on a
$400\times 400$ lattice.
Longitudinal (above) and transverse (below) 
correlations are shown at times
$180 (\bullet)$, $220 (\blacksquare)$,
$260 (\circ)$, $300 (\square)$, $340 (\triangle)$,
$380 (\diamondsuit)$, $420 (\star)$ and $460 (\ast)$. The solid lines
are the best OJK fits.}
\end{figure}

A theoretical prediction for the behavior 
of the pair correlation scaling function in models with 
non-conserved scalar field
is given by the OJK theory [\onlinecite{OJK}].
Monte Carlo data have been shown to be reproduced 
by the OJK theory better than by other theories after 
an appropriate rescaling of the temporal coordinate [\onlinecite{HB}].
The scaling function of  the OJK theory is given by
\begin{equation}
f(z)= \frac {2}{\pi}\sin^{-1}[\exp(-z^2/D)]
\label{eq:sf}
\end{equation}
where $ z=  r/t^{1/2}$ and $D=8(d-1)/d$. 
This gives the Porod linear behavior [\onlinecite{Po}]
at small z of the correlation function  $f(z) \sim 1 - \alpha z$
with $\alpha=2\sqrt{2}/(\pi\sqrt{D})$. Practically,
Monte Carlo and theoretical predictions can be compared 
imposing the same Porod behavior.
This procedure in our simulations gives two  different  
 $\alpha$ for the longitudinal and the transversal correlations,
$\alpha_{\ell}=0.383$ and $\alpha_t=0.414$ for $J_1=0.1$. The above discussed
symmetry corresponds to the fact that when $J_1=-0.1$
the best fit with the OJK function is given by $\alpha_{\ell}=0.406$ and 
$\alpha_t=0.376$.

The above results show that 
the OJK 
theory well describes the pair scaling function in anisotropic cases
if a free parameter is used to fit the data in the different directions.
Since the surface tension is the origin of the anisotropic
behavior, it is reasonable to 
check the validity of the OJK theory for anisotropic surface
tension models in the simplest case corresponding to
a field model 
with  anisotropic kinetic terms.
We consider the time dependent Ginzburg-Landau equation 
\begin{equation}
\frac {\partial\varphi}{\partial t} = 
B_x \frac{\partial^2\varphi}{\partial x^2} 
+ B_y \frac {\partial^2\varphi}{\partial y^2} -V'(\varphi) 
\label{eq:atdgl}
\end{equation}
where $V(\varphi)$ is the usual double-well potential. 
The rescaling $x \rightarrow x'=\sqrt{B_x} x$ and
$y \rightarrow y'=\sqrt{B_y} y$ would formally eliminate
the anisotropy and would give the usual spherically symmetric
OJK theory. However the correlations in terms of the original
space coordinates would have different Porod laws with
$\alpha_x/\alpha_y=\sqrt{B_y/B_x}$ showing an  anisotropic
behavior.
Of course, it is not possible
to simultaneously eliminate by a rescaling
both $B_x$ and $B_y$ when 
higher order derivative terms are present in the dynamical
equation,
so that the scaling function in a general anisotropic model
is expected to depend in some irriducible way 
on microscopic parameters. 
 
The anisotropic OJK behavior for the model (4)
can be tested by  studying  
the dynamical scaling of the Ising model with
different coupling $J_x$ and $J_y$ in the two lattice directions.
In Fig. 3 the scaled correlation data in the $x$ and the $y$
directions are compared with the OJK scaling function. 
The simulation are at finite temperature  with $J_x=2$
and $J_y=1$. Data for the usual
Ising model with $J_x=J_y=1$ are also plotted. The above
rescaling argument would suggest $\alpha_x/\alpha_y=\sqrt{J_y/J_x}$.
We measure a ratio $\alpha_y/\alpha_x=1.548$ which is not far 
from the expected ratio 1.414 and confirms
the fact that the OJK theory well reproduces the simulation
data in anisotropic cases.
The agreement with the theoretical expectation becomes stronger by  
decreasing the value of  the temperature. In the case $J_x/J_y=2$
and zero temperature we have measured  $\alpha_y/\alpha_x=1.4304$. 

To conclude, we have studied domain growth in a spin model
with four equivalent striped  ground states.
The pair correlation functions are different when measured
in the direction parallel and transversal to the stripes.
An explanation of this difference can be given on the basis
of the different interface energies between the four ground states.
Our results with single--spin--flip dynamics can be analyzed 
in the context of the Ohta-Jasnow-Kawasaki theory which is 
expected to  well reproduce scaling data in anisotropic surface tension
models. If other quantities such as 
the autocorrelation exponent or other correlation functions
here not considered  depend on the anisotropy of the system
is a matter for a future study.

\begin{figure}
\vskip -5mm
\inseps{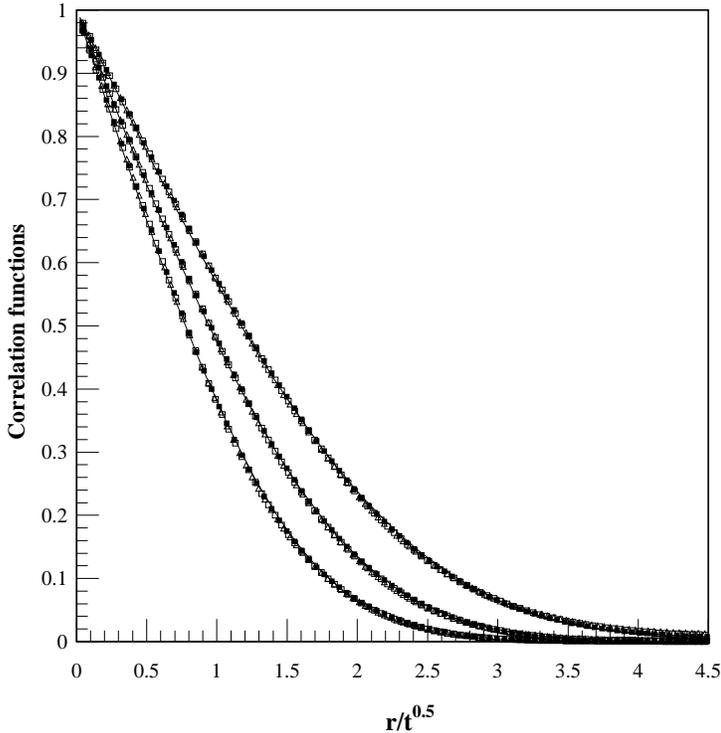}{0.6}
\vskip -10mm
\caption{Scaling collapse of the correlation functions of the Ising model at 
finite
temperature. Data were obtained averaging over $250$ stories in the
anisotropic case, $J_x=2$ and $J_y=1$, and over $447$ stories
in the isotropic case, $J_x=J_y=1$, on a $400\times 400$ square system.
>From above to below,
correlations along the x-direction in the anisotropic case,
correlations in the isotropic case, and correlation 
along the y-direction in the anisotropic case
 are shown at times
$350 (\blacksquare)$,
$450 (\square)$ and $500 (\triangle)$. The solid lines are the best
OJK fits.}
\end{figure}

~\\
\noindent ACKNOWLEDGEMENTS:\\
We thank Amos Maritan for helpful discussions.


\begin{references}
\bibitem{nota}
This procedure is basically the same used in [\onlinecite{Bind}]
where  $2 \times 2$ patterns were used.
\end{references}
\end{document}